\documentclass[useAMS,usenatbib]{mn2e}
\usepackage{graphicx}
\usepackage{txfonts}
\usepackage{subfigure}
\usepackage{natbib}
\title[From the core mass function to the stellar IMF]{Mapping the core mass function onto the stellar IMF:\\multiplicity matters}
\author[K. Holman, S. K. Walch, S. P. Goodwin and A. P. Whitworth]{K. Holman$^{1}$\thanks{E-mail: Katy.Holman@astro.cf.ac.uk (KH)}, S. K. Walch$^{1,2}$, S. P. Goodwin$^{3}$ and A. P. Whitworth$^{1}$\\
$^{1}$School of Physics and Astronomy, Cardiff University, Queens Buildings, The Parade, Cardiff, CF24 3AA, UK\\
$^{2}$Max-Planck-Institut f{\" u}r Astrophysik, Karl-Schwarzschild-Str. 1, Garching, D-85741, Germany\\
$^{3}$Department of Physics and Astronomy, University of Sheffield, Hicks Building, Housfield Road, Sheffield S3 7RH, UK}
\begin{document}

\date{Accepted . Received ; in original form }

\pagerange{\pageref{firstpage}--\pageref{lastpage}} \pubyear{2002}

\maketitle

\label{firstpage}

\begin{abstract}
Observations indicate that the central portions of the Present-Day Prestellar Core Mass Function (hereafter CMF) and the Stellar Initial Mass Function (hereafter IMF) both have approximately log-normal shapes, but that the CMF is displaced to higher mass than the IMF by a factor $F\!\sim\!4\!\pm\!1$. This has lead to suggestions that the shape of the IMF is directly inherited from the shape of the CMF -- and therefore, by implication, that there is a self-similar mapping from the CMF onto the IMF. If we assume a self-similar mapping, it follows (i) that $F\!=\!{\cal N}_{_{\rm O}}/\eta$, where $\eta$ is the mean fraction of a core's mass that ends up in stars, and ${\cal N}_{_{\rm O}}$ is the mean number of stars spawned by a single core; and (ii) that the stars spawned by a single core must have an approximately log-normal distribution of relative masses, with universal standard deviation $\sigma_{_{\rm O}}$. Observations can be expected to deliver ever more accurate estimates of $F$, but this still leaves a degeneracy between $\eta$ and ${\cal N}_{_{\rm O}}$; and $\;\sigma_{_{\rm O}}$ is also unconstrained by observation. Here we show that these parameters can be estimated by invoking binary statistics. Specifically, if (a) each core spawns one long-lived binary system, and (b) the probability that a star of mass $M$ is part of this long-lived binary is proportional to $M^{\alpha}$, current observations of the binary frequency as a function of primary mass, $b(M_{_1})$, and the distribution of mass ratios, $p_q$, strongly favour $\,\eta\!\sim\! 1.0\pm 0.3$,  $\,{\cal N}_{_{\rm O}}\!\sim\! 4.3\pm 0.4$, $\,\sigma_{_{\rm O}}\!\sim\! 0.3\pm 0.03\,$ and $\,\alpha\!\sim\! 0.9\pm 0.6$; $\;\eta\!>\!1$ just means that, between when its mass is measured and when it finishes spawning stars, a core accretes additional mass, for example from the filament in which it is embedded. If not all cores spawn a long-lived binary system, $db/dM_{_1}\!<\!0$, in strong disagreement with observation; conversely, if a core typically spawns more than one long-lived binary system, then ${\cal N}_{_{\rm O}}$ and $\eta$ have to be increased further. The mapping from CMF to IMF is not necessarily self-similar -- there are many possible motivations for a non self-similar mapping -- but if it is not, then the shape of the IMF cannot be inherited from the CMF. Given the limited observational constraints currently available and the ability of a self-similar mapping to satisfy them, the possibility that the shape of the IMF is inherited from the CMF cannot be ruled out at this juncture.
\end{abstract}


\section{Introduction} \label{SEC:INTRO}%

Understanding the processes that determine the IMF, and why these processes appear to vary little with environment and metallicity, is one of the main challenges in star formation \citep[e.g.][]{Elmegreen2008}. Recent observations of prestellar cores (i.e. the dense, gravitationally bound condensations in molecular clouds that are presumed to be destined to form individual stars or multiple systems) suggest that such cores have a mass function very similar in shape to the IMF, but shifted to higher masses by a factor of three to five \citep[e.g.][]{MotteetalCMF1998, Testi+SargentCMF1998, JohnstoneetalCMF2000, MotteetalCMF2001, JohnstoneetalCMF2001, StankeetalCMF2006, Enochetal2006, Johnstone+BallyCMF2006, YoungetalCMF2006, Nutter+WardTCMF2007, Alvesetal2007, Enochetal2008, SimpsonetalCMF2008, RathborneetalCMF2009, Veraetal2010}. The inference is that, in a statistical sense, there is a more-or-less self-similar mapping from prestellar cores to stars, and that the shape of the IMF is therefore simply inherited from the shape of the CMF. If true, this simply moves the problem to one of understanding the processes that determine the CMF, and why the outcome of these processes also varies little with environment and metallicity. In addition, we still need to understand how an individual core maps into an individual star or multiple system, and to what extent this process can really be viewed as statistically self-similar.

The IMF has been evaluated by \citet{KroupaStIMF2001} and \citet[][2005]{ChabrierStIMF2003}. \citeauthor{ChabrierStIMF2003} finds that the IMF is well fitted with a log-normal function merging into a power law at high masses. Theoretical models and simulations of turbulent fragmentation suggest that the CMF may also approximate to a log-normal function merging into a power law at high masses 
\citep{Padoan+NordlundCMF2002,PadoanetalCMF2007,Hennebelle+ChabrierCMF2008,Hennebelle+ChabrierCMF2009}.

However, these theories do not address the origins of stellar multiplicity. It is therefore timely to formulate the mapping between core mass and star mass using simple distribution functions, so that the additional constraints imposed by stellar multiplicity can be taken into account. It turns out that these additional constraints can be accomodated quite easily, but strongly favour a mapping in which each core typically spawns ${\cal N}_{_{\rm O}}\!\sim\!4$ stars, with quite high efficiency, $\eta\!\sim\! 1$; the individual stars spawned by a core have a log-normal mass distribution with standard deviation $\sigma_{_{\rm O}}\!\sim\!0.3$, and two of them end up in a long-lived binary system; the probability that a star with mass $M$ ends up in a long-lived binary system is approximately proportional to $M$.

\begin{table*}
\begin{center}
\caption{Input parameters regulating a single Monte Carlo integration, and the ranges of values admitted by the Markov chain. The prior for the Markov Chain is that, within these ranges, all values are equally probable. $M_{_{\rm C}}$ is the mass of a core, and $\{M_n\}_{n=1}^{n={\cal N}}\,$ are the masses of the stars formed from a single core.}
\begin{tabular}{clcc}\hline
$\;\;${\sc Parameter}$\;\;$ & {\sc Identity} & {\sc Minimum} & {\sc Maximum} \\\hline
$\mu_{_{\rm C}}$ & {\sc Arithmetic mean of} $\log_{_{10}}\!\left(M_{_{\rm C}}/{\rm M}_{_\odot}\right)$ & $-0.2$ & $+0.2$ \\
$\sigma_{_{\rm C}}$ & {\sc Standard Deviation of} $\log_{_{10}}\!\left(M_{_{\rm C}}/{\rm M}_{_\odot}\right)$ & $0.3$ & $0.7$ \\
$\eta$ & {\sc Mean Star Formation Efficiency in Core} $=\sum_{n=1}^{n={\cal N}}\left\{M_n\right\}/M_{_{\rm C}}\;\;\;$ & $0.0$ & $2.0$ \\
${\cal N}_{_{\rm O}}$ & {\sc Mean Number of Stars Formed in Core} & $1.0$ & $7.0$ \\ 
$\sigma_{_{\rm O}}$ & {\sc Standard Deviation of} $\log_{_{10}}\!\left(M_n/{\rm M}_{_\odot}\right)$ & $0.0$ & $0.5$ \\
$\alpha$ & {\sc Dynamical biasing parameter}, $\;d\ln(p_{_M})/d\ln(M)$ & $-2.0$ & $5.0$ \\\hline
\end{tabular}
\label{TAB:INPUT}
\end{center}
\end{table*}

In the interests of simplicity, we ignore the high-mass power-law parts of the mass functions, and concentrate on the log-normal parts, since these are the parts that are best constrained by observation, and they can be described with just two parameters: a logarithmic\footnote{Throughout, all logarithms are to base 10.} mean and standard deviation. Therefore our conclusions are most pertinent to the mass range where this log-normal form appears to be an acceptable approximation, say $0.03\;{\rm to}\;3\,{\rm M}_{_\odot}$. However, it should be noted that our conclusions are not significantly changed if the high-mass power-law tail is included; this simply makes the maths more laborious and less precise. For a detailed discussion of the IMF and the eight parameters that may be needed to describe it more completely, the reader is refered to \citet{BastianNetal2010}. We limit our consideration of multiplicity statistics to (i) the binary frequency as a function of primary mass, and (ii) the distribution of mass ratios (for systems with Sun-like and M-dwarf primaries), again because these appear to be the multiplicity statistics that are most robustly constrained by observation. For the purpose of this paper brown dwarfs are counted as stars.

In Section \ref{SEC:MOD} we present the definitions and assumptions underlying our model. In Section \ref{SEC:OBS} we present the observational data we will use to estimate the model parameters. In Section \ref{SEC:INF} we describe the consequences of the model, using simple arguments; this discussion pre-empts the results of the more rigorous statistical analysis that follows. In Section \ref{SEC:MOC} we describe how stellar statistics are evaluated for a particular model using Monte Carlo integration; and in Section \ref{SEC:FIT} we define the parameter we use to measure the quality of fit between a model and the observations. In Section \ref{SEC:MCRW} we describe the Markov Chain procedure for identifying the best-fit model parameters, and in Section \ref{SEC:RES} we present the results. In Section \ref{SEC:DIS}, we discuss the results and relate them to previous work, and in Section \ref{SEC:CON} we summarise our main conclusions.

\section{The Model} \label{SEC:MOD}%

\subsection{Assumptions}%

If one accepts that most stars are formed in cores \citep[see e.g.][]{Bressertetal10}, the model has only four assumptions.

{\sc Assumption I}. The central portions of the CMF and the IMF are both log-normal.

{\sc Assumption II}. The mapping between them is statistically self-similar, which means that the distribution of the {\it relative} masses of the stars spawned by a single core must also be log-normal.

{\sc Assumption III}. When a core forms more than one star, two of these stars end up in a binary system that is sufficiently long-lived to contribute to the statistics of binaries in the field. All the rest ultimately end up as singles.

{\sc Assumption IV}. The relative probability that a star with mass $M$ ends up in a long-lived binary system is proportional to $M^{\alpha}$.  

We note that these assumptions are not made because we believe they are necessarily true, but because they are simple, and because it turns out that they suffice to fit all the observational constraints that currently appear to be robust.

In addition, we note that the long-lived binary systems that contribute to the field statistics are probably not the only ones that form in a core-cluster, simply the ones that survive its dissolution and subsequent tidal perturbations \citep[e.g.][]{Kroupa1995}. There is evidence \citep[e.g.][]{Kohler2008,Chenetal2013} that the multiplicity is much higher than in the field for young stars in some star formation regions, and includes a significant proportion of higher-order multiples. However, by the time stars arrive in the field, many of these systems are likely to have been destroyed, and the wider systems will continue to suffer attrition due to stochastic tidal perturbations.

In other words, there are two very different timescales involved in the mapping. The mean number of stars spawned by a core (${\cal N}_{_{\rm O}}$) and the mean total mass of the stars spawned by a core (hence the efficiency, $\eta_{_{\rm O}}$) are -- ignoring stellar mass-loss, accretion, mass exchange and mergers -- determined by processes that terminate once the core disperses, after at most a few mega-years. In contrast, the binary statistics are never completely settled. They evolve most rapidly during the birth throes of the core-cluster (the ${\cal N}_{_{\rm O}}$ stars formed from a single core) and during its dispersion, but they then evolve further due to interactions with other stars in the same large-scale cluster (here presumed to be an ensemble of stars formed from an ensemble of cores), and they continue to evolve, after the large-scale cluster dissolves, due to interactions with the ever changing background gravitational field (e.g. tidal perturbations from passing stars and molecular clouds). However these latter perturbations are rare, and given that the typical field star has been in the field for many giga-years, its binary statistics should by now be well defined. Our model does not concern itself with the details of the dynamical evolution of the stars spawned by a single core; it simply focuses on the properties of systems that survive to populate the field, posits that each core typically spawns just one such system, and shows that the observed binary statistics are reproduced well if this system tends to comprise two of the more massive stars spawned by the core. Other binary systems, and higher multiples, are spawned by a core, but we presume that they are disrupted on a timescale $\la\!1\,{\rm Gyr}$. One would expect the binary systems surviving in the field to be on average more massive and closer than the ones that have been disrupted.

\subsection{Input parameters}%

Table \ref{TAB:INPUT} summarises the six model input parameters, viz. the logarithmic mean, $\mu_{_{\rm C}}$, and standard deviation, $\sigma_{_{\rm C}}$, of the CMF; the efficiency, $\eta$, i.e. the fraction of a core's mass that is converted into stars; the mean number of stars, ${\cal N}_{_{\rm O}}$, spawned by a single core; the logarithmic standard deviation, $\sigma_{_{\rm O}}$, of the relative masses of the stars spawned by a core; and the dynamical biasing parameter, $\alpha$. There are direct observational constraints on $\mu_{_{\rm C}}$ and $\sigma_{_{\rm C}}$, but not, as yet, on $\eta$, ${\cal N}_{_{\rm O}}$, $\sigma_{_{\rm O}}$ and $\alpha$.

We note that values of $\eta$ greater than unity are admissible, because, between the time when the mass of a core is estimated and added to the CMF, and the time when its star formation is complete, the core can, and almost certainly does, grow in mass, for example by accretion along the filament in which it is embedded \citep[e.g.][]{SmithR2011}. By the same token it is not necessary that all the stars spawned by a core form simultaneously. Indeed, numerical simulations suggest that some of the stars spawned by a core start to condense out of the filamentary material accreting onto the core, and may only reach the core as it starts to disintegrate \citep[e.g.][]{Bate2012, Girichidis2012}

In addition, non-integer values of ${\cal N}_{_{\rm O}}$ are admissible. In such cases, we adopt the simple device of dividing cores between the integer values that bracket ${\cal N}_{_{\rm O}}$. Thus, for example, ${\cal N}_{_{\rm O}}=2.2$ means that $80\%$ of cores have ${\cal N}=2$ and $20\%$ have ${\cal N}=3$.

Apart from this device, we do not allow any variance in the input parameters, because to do so introduces extra input parameters, but does not significantly improve, or even alter, the fits obtained.

\subsection{Output parameters}%

Given the four assumptions listed above, and values for the six input parameters, we can predict the IMF (which, being log-normal, is characterised by a logarithmic mean, $\mu_{_{\rm S}}$, and a logarithmic standard deviation, $\sigma_{_{\rm S}}$), the binary frequency as a function of primary mass, $b(M_{_1})$, and the distributions of mass ratio for systems with Sun-like and M-dwarf primaries, $p_q(M_{_1})$. Our objective is to use observations of these output parameters ($\mu_{_{\rm S}}$, $\sigma_{_{\rm S}}$, $b(M_{_1})$, $p_q(M_{_1})$) to constrain the model input parameters ($\mu_{_{\rm C}}$, $\sigma_{_{\rm C}}$, $\eta$, ${\cal N}_{_{\rm O}}$, $\sigma_{_{\rm O}}$, $\alpha$). 

\begin{table*}
\begin{center}
\caption{Output parameters characterising the observed IMF and binary statistics (two lefthand columns), and parameters regulating the quality of the fit of a model to the observations (three righthand columns). Column 1 gives the name of the parameter in the model, and Column 2 its identity. Column 3 gives the observed {\bf V}alue ($V$) of this parameter, and Column 4 its {\bf U}ncertainty ($U$). Column 5 gives the {\bf W}eight ($W$) accorded to fitting the observed value. $M_{_{\rm S}}$ is the mass of a star from the whole ensemble of stars formed in a single Monte Carlo integration. The sources for the observational data are given in Section \ref{SEC:OBS}.}
\begin{tabular}{cllll}\hline
{\sc Parameter} & {\sc Identity} & {\sc Observed Value} & {\sc Uncertainty}\hspace{0.45cm} & {\sc Weight} \\\hline
$\mu_{_{\rm S}}$ & {\sc Mean of} $\log_{_{10}}\!\left(M_{_{\rm S}}/{\rm M}_{_\odot}\right)$ & $V_{\mu_{_{\rm S}}}=-0.70$ & $U_{\mu_{_{\rm S}}}=0.10$ & $W_{\mu_{_{\rm S}}}=1/4$ \\
$\sigma_{_{\rm S}}$ & {\sc Standard Deviation of} $\log_{_{10}}\!\left(M_{_{\rm S}}/{\rm M}_{_\odot}\right)\hspace{0.7cm}$ & $V_{\sigma_{_{\rm S}}}=0.55$ &$U_{\sigma_{_{\rm S}}}=0.05$ & $W_{\sigma_{_{\rm S}}}=1/4$ \\\\
$b_1$ & {\sc Multiplicity Frequency in} $\,(0.05,0.10)\,{\rm M}_{_\odot}\;\;\;\;\;\;\;$ & $V_{b_1}=0.20$ & $U_{b_1}=0.15$ & $W_{b_1}=1/16$ \\
$b_2$ & {\sc Multiplicity Frequency in}  $\,(0.05,0.17)\,{\rm M}_{_\odot}$ & $V_{b_2}=0.26$ & $U_{b_2}=0.10$ & $W_{b_2}=1/16$ \\
$b_3$ & {\sc Multiplicity Frequency in}  $\,(0.15,0.60)\,{\rm M}_{_\odot}$   & $V_{b_3}=0.34$ & $U_{b_3}=0.04$ & $W_{b_3}=1/16$ \\
$b_4$ & {\sc Multiplicity Frequency in}  $\,(0.8,1.2)\,{\rm M}_{_\odot}$   & $V_{b_4}=0.45$ & $U_{b_4}=0.03$ & $W_{b_4}=1/16$ \\
$b_5$ & {\sc Multiplicity Frequency in}  $\,(3,50)\,{\rm M}_{_\odot}$      & $V_{b_5}=0.70$ & $U_{b_5}=0.10$ & $W_{b_5}=0$ \\
$b_6$ & {\sc Multiplicity Frequency in}  $\,(20,70)\,{\rm M}_{_\odot}$     & $V_{b_6}=0.85$ & $U_{b_6}=0.10$ & $W_{b_6}=0$ \\\\
$p_{3,\ell}$ & {\sc Fraction of systems from primary-mass bin 3 in mass-ratio bin} $\ell\;\;(\ell\!=\!1\,{\rm to}\,5)$\hspace{0.6cm} 
   & $V_{p_{3,1}}=0.20$ & $U_{p_{3,1}}=0.05$ & $W_{p_{3,1}}=1/40$ \\
 & & $V_{p_{3,2}}=0.20$ & $U_{p_{3,2}}=0.05$ & $W_{p_{3,2}}=1/40$ \\
 & & $V_{p_{3,3}}=0.20$ & $U_{p_{3,3}}=0.05$ & $W_{p_{3,3}}=1/40$ \\
 & & $V_{p_{3,4}}=0.20$ & $U_{p_{3,4}}=0.05$ & $W_{p_{3,4}}=1/40$ \\
 & & $V_{p_{3,5}}=0.20$ & $U_{p_{3,5}}=0.05$ & $W_{p_{3,5}}=1/40$ \\\\
$p_{4,\ell}$ & {\sc Fraction of systems from primary-mass bin 4 in mass-ratio bin} $\ell\;\;(\ell\!=\!1\,{\rm to}\,5)$
   & $V_{p_{4,1}}=0.10$ & $U_{p_{4,1}}=0.03$ & $W_{p_{4,1}}=1/40$ \\
 & & $V_{p_{4,2}}=0.25$ & $U_{p_{4,2}}=0.05$ & $W_{p_{4,2}}=1/40$ \\
 & & $V_{p_{4,3}}=0.21$ & $U_{p_{4,3}}=0.05$ & $W_{p_{4,3}}=1/40$ \\
 & & $V_{p_{4,4}}=0.19$ & $U_{p_{4,4}}=0.04$ & $W_{p_{4,4}}=1/40$ \\
 & & $V_{p_{4,5}}=0.25$ & $U_{p_{4,5}}=0.05$ & $W_{p_{4,5}}=1/40$ \\\hline
\end{tabular}
\label{TAB:STELSTAT}
\end{center}
\end{table*}

\section{Observational data}\label{SEC:OBS}%

Table \ref{TAB:STELSTAT} summarises the expectation values, $V_X$, uncertainties, $U_X$, and weights, $W_X$, accorded to the different observational parameters, $X$, that the model seeks to predict. The weights determine the influence that different observed quantities exert on the overall quality of fit of a model (see Section \ref{SEC:FIT}), and by design they add up to unity. 

For the mean and standard deviation of the IMF, $\mu_{_{\rm S}}$ and $\sigma_{_{\rm S}}$, we use values informed by \citet{Chabrier2005}, and since these two quantities appear to be quite well constrained by observation, we give them both a high weight, $W_{\mu_{_{\rm S}}}=W_{\sigma_{_{\rm S}}}=1/4$. 

For the binary frequencies we consider six primary-mass bins. Bin $m\!=\!1$ (the lowest mass bin) represents the results of \citet{Closeetal2003}; Bin 2, those of \citet{Basri+Reiners2006}; Bin 3, those of \citet{Janson2012}; Bin 4, those of \citet{Raghavan2010}; Bin 5, those of \citet{Preibischetal1999}; and Bin 6, those of \citet{MasonBetal1998}. For evaluating the quality of the fit, we give the first four bins equal weights, $W_{b_i}=1/16,\;i\!=\!1\,{\rm to}\,4$, so that their combined weight is $1/4$. The last two bins are given zero weight, because the stars in these bins are not strictly field stars.\footnote{The last two bins concern binaries with relatively high-mass short-lived primaries in the Orion Nebula Cluster \citep{Preibischetal1999} and a mixture of systems in clusters, associations and the field \citep{MasonBetal1998}.} Therefore these two bins should not influence the choice of best-fit model. They are included because -- notwithstanding -- the predictions of the best-fit model agree with them well (see Fig. \ref{FIG:BINFREQ}).

For the distribution of mass ratios, $q$, we consider only primary-mass bins $m\!=\!3\;{\rm and}\;4$, since these are the ones with relatively robust mass-ratio statistics \citep{Raghavan2010,Reggiani2011,Janson2012}. In both primary-mass bins, the distribution of mass ratios appears to be flat \citep{Reggiani2011}. We follow convention by allocating the mass-ratios to five equal bins, $\ell\!=\!1\,{\rm to}\,5$, so that bin $\ell$ accommodates values in the range $0.2(\ell\!-\!1)\!<\!q\!\leq\!0.2\ell$. For primary-mass bin 3, \citet{Janson2012} conclude that, when allowance is made for selection effects, the distribution of mass ratios is flat, and therefore we simply set all the expectation values to $V_{p_{3,\ell}}\!=\!0.20$, and all the uncertainties to $U_{p_{3,\ell}}\!=\!0.05$. For primary-mass bin 4, we adopt expectation values and Poisson uncertainties from \citet{Raghavan2010}. For all ten primary-mass/mass-ratio bins we allocate $W_{p_{m,\ell}}=1/40$, so that their combined weight is $1/4$.

\section{Simple inferences} \label{SEC:INF}%

In Section \ref{SEC:RES} we present the results of a Markov Chain Monte Carlo analysis. Here we present simple arguments to preempt the main results of that analysis.

\subsection{The shift between the IMF and the CMF}\label{SEC:SHIFT}%

The mean mass of the stars that form from a given core are related to the mass of the core by the efficiency, $\eta$ (the fraction of the core's mass that ends up in stars), divided by the number of stars formed from the core ${\cal N}_{_{\rm O}}$. Hence the factor by which the peak of the CMF exceeds the peak of the IMF is given by
\begin{eqnarray}\label{EQN:SHIFT}
F&\equiv&10^{\,(\mu_{_{\rm C}}-\mu_{_{\rm S}})}\;\,=\;\,\frac{{\cal N}_{_{\rm O}}}{\eta}\,.
\end{eqnarray}
If we adopt $\mu_{_{\rm S}}\!=\!-0.6\!\pm\!0.05$ \citep[from][]{ChabrierStIMF2003}, and $\mu_{_{\rm C}}\!=\!0.0\!\pm\!0.1$ \citep[from, e.g.,][]{Enochetal2006, YoungetalCMF2006, Enochetal2008, Veraetal2010}, we have $F\simeq 4\pm 1\,$, whence
\begin{eqnarray}\label{EQN:DEGEN}
{\cal N}_{_{\rm O}}\;\,=\;\,F\,\eta&\simeq&(4\pm 1)\;\eta\,.
\end{eqnarray}

\subsection{Raising the degeneracy between ${\cal N}_{_{\rm O}}$ and $\eta$}\label{SEC:GED}%

The degeneracy between ${\cal N}_{_{\rm O}}$ and $\eta$ can be raised by considering the binary statistics. Two essential features of the binary statistics in the field are that -- very roughly -- the number of single-star systems is comparable with, but somewhat larger than, the number of binary systems, {\it and} the binary frequency is an increasing function of primary mass ($db/dM_{_1}>0$). The influence of these constraints can be understood with the following {\it Gedankenexperiment}. Suppose (purely for the sake of argument, and averaged over all masses) that 60\% of systems are single and 40\% are binary. This can be achieved in two ways.

\begin{itemize}

\item{${\cal N}_{_{\rm O}}=1.4$. In this case, $60\%$ of cores have ${\cal N}=1$ and spawn singles, whilst $40\%$ of cores have ${\cal N}=2$ and spawn binaries. This gives $0.26\la\eta\la 0.44$. However, it means that the components of binary systems are on average less massive than single stars, and therefore the binary fraction is a decreasing function of primary mass, which is the opposite of what is observed.}

\item{${\cal N}_{_{\rm O}}=3.5$. In this case, each core spawns a binary system, but $50\%$ have ${\cal N}\!=\!3$ so they spawn one extra single star, and the remaining $50\%$ have ${\cal N}\!=\!4$ and therefore spawn two extra single stars. This gives $0.7\la\eta\la 1$. Moreover, provided $\alpha\!>\!0$, the components of binary systems are now, on average, more massive than the single stars, and consequently the binary fraction is an increasing function of primary mass, as observed.}

\end{itemize}

There is therefore a strong preference for the larger value of ${\cal N}_{_{\rm O}}$, to ensure that $db/dM_{_1}\!>\!0$.

\subsection{Standard deviation of the relative masses of the stars spawned by a single core}%

Since the mapping of the CMF onto the IMF involves the convolution of a log-normal CMF with a log-normal distribution of relative stellar masses, the logarithmic standard deviation of the IMF, $\sigma_{_{\rm S}}$, is obtained by adding the logarithmic standard deviation of the CMF, $\sigma_{_{\rm C}}$, and the logarithmic standard deviation of the relative stellar masses, $\sigma_{_{\rm O}}$, in quadrature,
\begin{eqnarray}\label{EQN:ADDSIG}
\sigma_{_{\rm S}}^2&=&\sigma_{_{\rm C}}^2\;+\;\sigma_{_{\rm O}}^2\,.
\end{eqnarray}
A corollary of Eqn. (\ref{EQN:ADDSIG}) is that -- for a self-similar mapping -- the logarithmic standard deviation of the IMF cannot be smaller than the logarithmic standard deviation of the CMF,
\begin{eqnarray}
\sigma_{_{\rm S}}&\geq&\sigma_{_{\rm C}}\,.
\end{eqnarray}
In interpreting this inequality, one must recognise that the log-normal CMF we are discussing here is one that represents a very large region embracing a representative ensemble of star formation regions; the log-normal CMFs inferred for individual star formation regions can -- and apparently do -- have a range of means and logarithmic standard deviations, but together they cannot have a logarithmic standard deviation greater than that of the IMF and still admit a self-similar mapping. Since observations suggest $\sigma_{_{\rm C}}\!\sim\!\sigma_{_{\rm S}}$, this in turn implies that $\sigma_{_{\rm O}}$ cannot be {\it very large}.

\subsection{Mass ratios}%

Observations \citep{Raghavan2010,Reggiani2011,Janson2012} suggest that the distributions of mass ratio for binary systems having Sun-like and M-dwarf primaries are both flat. In our model, this means firstly that $\sigma_{_{\rm O}}$ can not be {\it very small},\footnote{It turns out that finding a value of $\sigma_{_{\rm O}}$ that is both small enough to satisfy Eqn. \ref{EQN:ADDSIG}, and large enough to deliver low-$q$ binaries, is the hardest constraint for the model to satisfy.} otherwise the range of stellar masses formed in a single core would be too narrow to produce low-$q$ systems; and secondly that $\alpha$ can not be too large, otherwise the low-mass stars would have little chance of pairing up with the high-mass ones to prouce low-$q$ systems.

\section{Monte Carlo integration} \label{SEC:MOC}%

For a single model (i.e. a fixed combination of the input parameters, $\mu_{_{\rm C}},\,\sigma_{_{\rm C}},\,\eta,\,{\cal N}_{_{\rm O}},\,\sigma_{_{\rm O}},\,\alpha$), we evaluate the stellar statistics as follows.

First, a core mass, $M_{_{\rm C}}$, is obtained by generating a Gaussian random deviate, ${\cal G}$, on $(-\infty,+\infty)$, and setting
\begin{eqnarray}
M_{_{\rm C}}&=&10^{(\mu_{_{\rm C}}+{\cal G}\sigma_{_{\rm C}})}\,{\rm M}_{_\odot}\,.
\end{eqnarray}
Next, if ${\cal N}_{_{\rm O}}$ is non-integer, a value for ${\cal N}$ is obtained by generating a linear random deviate, ${\cal L}$, on $(0,1)$, and putting
\begin{eqnarray}
{\cal N}&=&\left\{\begin{array}{ll}
{\rm INT}({\cal N}_{_{\rm O}}),\hspace{0.4cm}&{\rm when}\;{\cal L}\geq{\cal N}_{_{\rm O}}-{\rm INT}({\cal N}_{_{\rm O}});\\
{\rm INT}({\cal N}_{_{\rm O}})+1,&{\rm when}\;{\cal L}<{\cal N}_{_{\rm O}}-{\rm INT}({\cal N}_{_{\rm O}}).\\
\end{array}\right.
\end{eqnarray}
Otherwise ${\cal N}\!=\!{\cal N}_{_{\rm O}}$. Then the masses of the ${\cal N}$ stars spawned by this core can be obtained by generating Gaussian random deviates, ${\cal G}$, on $(-\infty,+\infty)$, and computing
\begin{eqnarray}
M_{_{\rm S}}&=&\frac{M_{_{\rm C}}\eta}{\cal N}\,10^{{\cal G}\sigma_{_{\rm O}}}\,.
\end{eqnarray}
If ${\cal N}\!\geq\!2$, the integrated probability of each possible pairing of these stars (star $n$ with star $n'$) is computed,
\begin{eqnarray}
P_{n,n'}&=&\frac{\sum\limits_{\nu=1}^{\nu=n}\,\sum\limits_{\nu'=\nu+1}^{\nu'=n'}\,\left\{M_\nu^{\alpha}\,M_{\nu'}^{\alpha}\right\}}{\sum\limits_{\nu=1}^{\nu={\cal N}-1}\,\sum\limits_{\nu'=\nu+1}^{\nu'={\cal N}}\,\left\{M_\nu^{\alpha}\,M_{\nu'}^{\alpha}\right\}}\,.
\end{eqnarray}
Finally, a linear random variate, ${\cal L}$, on $(0,1)$, is generated, and the pairing whose integrated probability is just above ${\cal L}$ is selected.

This is repeated until a total of $10^7$ stars has been created. Then the mean and standard deviation of the IMF, $\mu_{_{\rm S}}$ and $\sigma_{_{\rm S}}$, are computed (using the logarithms of the stellar masses). For each star that falls in one of the mass bins defined in Table \ref{TAB:STELSTAT}, we note (i) whether it is the primary in a binary system, the secondary in a binary system, or a single star; and, if it is a primary, we also note which mass-ratio bin the binary falls in. If mass bin $m$ contains $P_m$ primaries and $S_m$ singles, the corresponding binary frequency\footnote{We refer the reader to \citet{Reipurth+Zinnecker1993} for a discussion of different measures of multiplicity and their various merits. The one defined in Eqn. (\ref{EQN:BINFREQ}) is in effect the multiplicity frequency, but we refer to it as the binary frequency because we are only considering binaries. As pointed out by \citet{Hubber+Whitworth2005}, the multiplicity frequency has the nice property that it is insensitive to whether a binary system is actually a higher-order multiple. We note parenthetically that there are in general other stars in each mass bin that are secondaries, but these do not explicitly affect the calculation of the $b_m$.} is
\begin{eqnarray}\label{EQN:BINFREQ}
b_m&=&\frac{P_m}{P_m+S_m}\,.
\end{eqnarray}
If mass-ratio bin $\ell$ of mass bin $m$ contains $C_{m\ell}$ systems, the corresponding mass-ratio probability is
\begin{eqnarray}
p_{m,\ell}&=&\frac{C_{m\ell}}{P_m}\,.
\end{eqnarray}
The model can then be compared with the observational data.

\begin{figure*}
\begin{center}
\includegraphics[width=0.333\textwidth,angle=270]{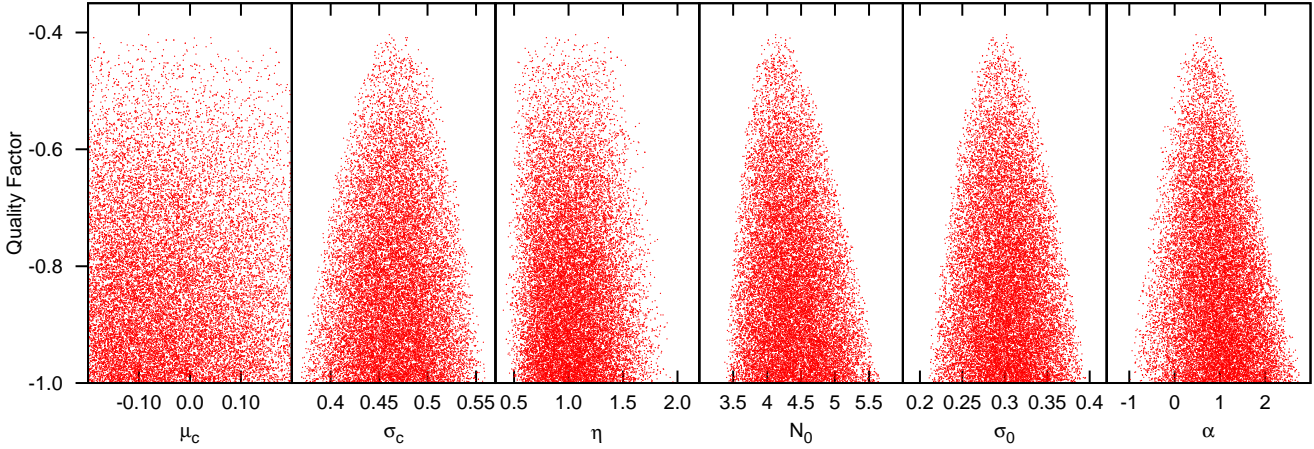}
\caption{The ${\cal Q}$-values for all models along the Markov Chain that have ${\cal Q}>-1$, plotted against $\mu_{_{\rm C}}$, $\sigma_{_{\rm C}}$, $\eta$, ${\cal N}_{_{\rm O}}$, $\sigma_{_{\rm O}}$ and $\alpha$.}
\label{FIG:MARKOVDOTS}
\end{center}
\end{figure*}

\section{Quality of fit}\label{SEC:FIT}%

For each model (i.e. each Monte Carlo integration with a given set of input parameters, $\mu_{_{\rm C}},\,\sigma_{_{\rm C}},\,\eta,\,{\cal N}_{_{\rm O}},\,\sigma_{_{\rm O}},\,\alpha$), the quality of fit, ${\cal Q}$, is given by a sum of terms,
\begin{eqnarray}\label{EQN:QGEN}
\Delta{\cal Q}_X&=&-\;\frac{W_X\,(X-V_X)^2}{U_X^2}\,,
\end{eqnarray}
representing how well the model prediction for output parameter $X\;\;(\equiv\mu_{_{\rm S}},\,\sigma_{_{\rm S}},\,b_m\,[{\rm for}\,m\!=\!1,\,2,\,3,\,4],\,p_{m\ell}\,[{\rm for}\,m\!=\!3,\,4;\,\ell\!=\!1,\,2,\,3,\,4,\,5]$) matches with the observational constraints (see Table \ref{TAB:STELSTAT}). The overall quality of fit for a given model is then
\begin{eqnarray}\nonumber
{\cal Q}(\mu_{_{\rm C}},\sigma_{_{\rm C}},\eta,{\cal N}_{_{\rm O}},\sigma_{_{\rm O}},\alpha)=
-\,\frac{W_{\mu_{_{\rm S}}}(\mu_{_{\rm S}}-V_{\mu_{_{\rm S}}})^2}{U_{\mu_{_{\rm St}}}^2}\hspace{2.1cm}&&\\\nonumber
-\,\frac{W_{\sigma_{_{\rm S}}}(\sigma_{_{\rm S}}-V_{\sigma_{_{\rm S}}})^2}{U_{\sigma_{_{\rm S}}}^2}\hspace{2.0cm}&&\\\nonumber
-\,\sum\limits_{m=1}^{m=4}\left\{\frac{W_{b_m}(b_m-V_{b_m})^2}{U_{b_m}^2}\right\}\hspace{1.2cm}&&\\\label{EQN:QUALITY}
-\,\sum\limits_{m=3}^{m=4}\left\{\sum\limits_{\ell=1}^{\ell=5}\left\{\frac{W_{p_{m,\ell}}(p_{m,\ell}-V_{p_{m,\ell}})^2}{U_{p_{m,\ell}}^2}\right\}\right\}\!.
\end{eqnarray}
The first two terms on the righthand side of Eqn. (\ref{EQN:QUALITY}) measure the ability of the model to reproduce the observed IMF (with an overall weighting of $50\%$); the third term (involving a single summation) measures its ability to reproduce the observed binary frequency as a function of primary mass (with an overall weighting of $25\%$); and the fourth term (involving a double summation) measures its ability to reproduce the distributions of mass ratio for systems having Sun-like and M-dwarf primaries (with an overall weighting of $25\%$). A notionally perfect fit corresponds to ${\cal Q}=0$, and $|\,{\cal Q}\,|$ can be interpreted as the number of standard deviations by which the model departs from a perfect fit.

\section{Markov Chain}\label{SEC:MCRW}%

\subsection{Range of $\mu_{_{\rm C}}$ and $\sigma_{_{\rm C}}$}%

{\sc HERSCHEL} has allowed much more robust evaluations of the CMF. For example, \citet{Veraetal2010} obtain $(\mu_{_{\rm C}},\sigma_{_{\rm C}})\!=\!(-0.22,0.42)\;{\rm and}\;(-0.05,0.30)$ in -- respectively -- the entire Aquila field and the main Aquila subfield. Previously, \citet{Enochetal2006} have estimated $(\mu_{_{\rm C}},\sigma_{_{\rm C}})\!=\!(-0.05\!\pm\!0.25,0.50\!\pm\!0.10)$ in Perseus; \citet{YoungetalCMF2006} have estimated $(\mu_{_{\rm C}},\sigma_{_{\rm C}})\!=\!0.3\!\pm\!0.7,0.5\!\pm\!0.4)$ in Ophiuchus; and \citet{Enochetal2008} have estimated $(\mu_{_{\rm C}},\sigma_{_{\rm C}})\!=\!0.00\!\pm\!0.04,0.30\!\pm\!0.03)$ for an ensemble of cores from Perseus, Serpens and Ophiuchus.

However, all these evaluations are convolved with a number of uncertainties. In particular, the use of greybody fits to estimate mean dust temperatures, the mass opacity coefficients needed to convert fluxes into masses, and the distances assumed for the star-formation regions, all introduce uncertainty into the derived masses, and hence into the $\mu_{_{\rm C}}$-values. $\;\;\sigma_{_{\rm C}}$-values may be somewhat less susceptible to these factors, but are affected by the fact that the cores on the low-mass side of the log-normal tend to be close to the completeness limit. Furthermore, we are here concerned with the values of $\,\mu_{_{\rm C}}\,$ and $\,\sigma_{_{\rm C}}\,$ for the totality of all star forming cores, rather than those for a single region.

To keep the problem tractable, we restrict the Markov Chain to values of  $\mu_{_{\rm C}}$ in the range $-0.2\!<\!\mu_{_{\rm C}}\!<\!+0.2$. We discuss the consequences of taking $\mu_{_{\rm C}}$-values outside this range, in Section \ref{SEC:RES}. For $\sigma_{_{\rm C}}$ we restrict the Markov Chain to values of  $\sigma_{_{\rm C}}$ in the range $0.3\!<\!\sigma_{_{\rm C}}\!<\!0.7$. This choice is informed by the range of observationally inferred values, and by the fact that $\sigma_{_{\rm C}}$ cannot exceed $\sigma_{_{\rm S}}$.

\subsection{Range of $\eta$ and ${\cal N}_{_{\rm O}}$}%

We restrict the Markov Chain to values of $\eta$ in the range $0\!<\!\eta\!<\!2$, and values of ${\cal N}_{_{\rm O}}$ in the range $1\!\leq\!{\cal N}_{_{\rm O}}\!\leq\!7$. Evidently, if a core accretes very rapidly on the way to forming stars, higher $\eta$ values are possible, but this turns out to be unlikely. The arguments presented in Section \ref{SEC:GED} suggest that higher values of ${\cal N}_{_{\rm O}}$ are inadmissible -- unless each core spawns more than one long-lived binary, {\it and} the efficiency is increased still further (see Section \ref{SEC:SHIFT}).

\subsection{Range of $\sigma_{_{\rm O}}$ and $\alpha$}%

We restrict the Markov Chain to values of $\sigma_{_{\rm O}}$ in the range $0\!<\!\sigma_{_{\rm O}}\!<\!0.5$, on the grounds that $\sigma_{_{\rm O}}$ has to be smaller than $\sigma_{_{\rm S}}$, and is probably also smaller than $\sigma_{_{\rm C}}$.

We restrict the Markov Chain to values of $\alpha$ in the range $-2\!<\!\alpha\!<\!5$. This choice is informed by numerical work on the dissolution of small-$N$ clusters \citep[e.g.][and references therein]{VanAlbada68a,VanAlbada68b,McDonald+Clarke93,Sterzik+Durisen98,Hubber+Whitworth2005}, which suggests that, if the dissolution of a core-cluster involves pure gravitational interaction between the stars, a single long-lived binary is the most likely outcome and it usually  comprises the two most massive stars, which implies $\alpha\!\gg\!1$. Conversely, if there is dissipation -- for example, because the stars are attended by massive discs \citep{McDonald+Clarke95} -- other pairings become more likely, which implies a smaller $\alpha$ value. Flat mass-ratio distributions translate into a preference for small $\alpha$.

\begin{figure*}
\begin{center}
\includegraphics[width=0.90\textwidth,angle=00]{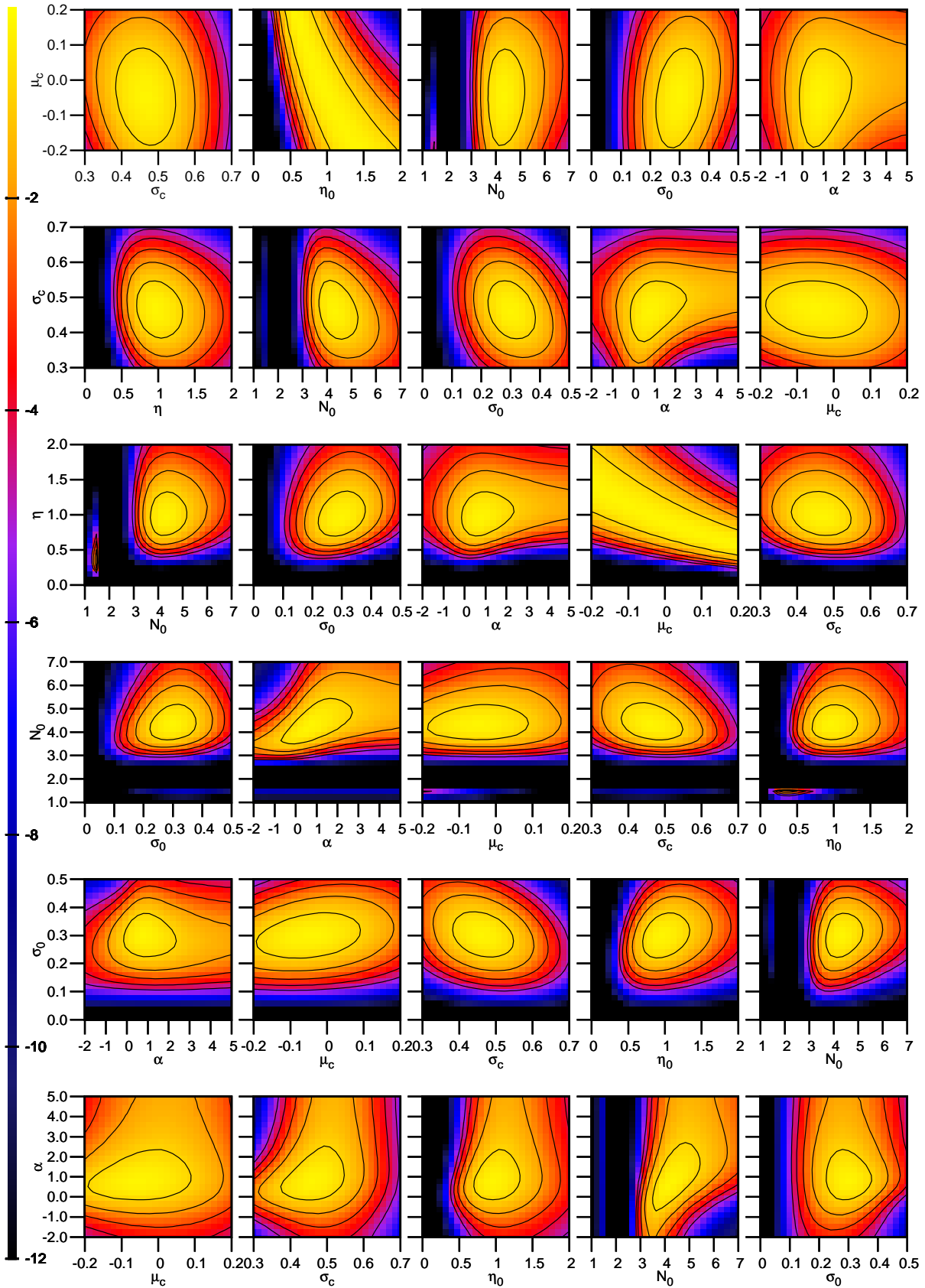}
\caption{Iso-${\cal Q}$ plots on principal planes through the best=fit model. On each row the ordinate (vertical axis) is the same model input parameter, from top to bottom in the order $\mu_{_{\rm C}}$, $\sigma_{_{\rm C}}$, $\eta$, ${\cal N}_{_{\rm O}}$, $\sigma_{_{\rm O}}$ and $\alpha$. Along each row the abscissa (horizontal axis) cycles through the remaining model input parameters, in the same order. By scanning along a row one can see both which parameters are tightly constrained by the model, and which parameters are correlated. The false colour encodes the value of ${\cal Q}$ (see bar on right of plot), and the contours correspond to ${\cal Q}\!+\!1,\,2,\,3,\,4,\,5$.}
\label{FIG:MONTAGE}
\end{center}
\end{figure*}

\subsection{Markov Chain}%

The ranges detailed above define the input parameter space, and our prior is that all values in these ranges are equally probable . The Markov Chain then starts at an arbitrary point in this space, and makes a biased random walk around the space. The components of a step are generated from Gaussian distributions. A step is always taken if $\Delta{\cal Q}={\cal Q}_{_{\rm NEW}}-{\cal Q}_{_{\rm OLD}}>0$ (i.e. if it results in an improvement to the fit). If $\Delta{\cal Q}<0$, the code generates a linear random deviate, ${\cal L}$, on $(0,1)$, and only takes the step if $\Delta{\cal Q}>\ln({\cal L})$ (i.e. steps that produce a deterioration in the fit are less likely to be taken the larger the deterioration). The size of a step is scaled so that roughly half of all putative steps are not taken.

\section{Results}\label{SEC:RES}%

From the Markov Chain, there is a single well defined ${\cal Q}$ peak in the parameter space explored, and the best fit is obtained with
\begin{eqnarray}
\mu_{_{\rm C}}&=&-0.03\,\pm\,0.10\,,\\
\sigma_{_{\rm C}}&=&0.47\,\pm\,0.04\,,\\
\eta&=&1.01\,\pm\,0.27\,,\\
{\cal N}_{_{\rm O}}&=&4.34\,\pm\,0.43\,,\\
\sigma_{_{\rm O}}&=&0.30\,\pm\,0.03\,,\\
\alpha&=&0.87\,\pm\,0.64\,,\\
{\cal Q}&=&-0.33\,,
\end{eqnarray}
i.e. $0.33\sigma$ overall difference between the model and the observations.

The parameters of the CMF ($\mu_{_{\rm C}},\sigma_{_{\rm C}}$) are compatible with those obtained from observation, although $\mu_{_{\rm C}}$ has a rather large uncertainty, and we return to this point below.

The efficiency ($\eta$) is much higher than the values normally estimated \citep[e.g.][]{Alvesetal2007}. $\eta$ is also only just compatibleat the high end of the range calculated theoretically by \citet{Matzner+McKee2000}, but in their model these high values arise in cores that are intrinsically flattened (so that outflows can escape without sweeping up much core mass), rather than as a consequence of forming many stars. High notional efficiencies may be an indication that cores grow in mass whilst they collapse and fragment to form stars \citep[e.g.][]{SmithR2011}.

The mean number of stars formed from a single core (${\cal N}_{_{\rm O}}$) is also higher than the values normally invoked. Mathematically this follows from the large $\eta$ (see Eqn. \ref{EQN:SHIFT}), but physically it also derives -- inevitably, in a self-similar mapping -- from the need to form binaries with a frequency that increases with primary mass (see discussion in Section \ref{SEC:GED}).

The spread of stellar masses from a single core-cluster ($\sigma_{_{\rm O}}=0.29\pm0.07$) is such that, if the stars are paired randomly, between 33 and $56\%$ of the resulting systems have mass ratio below 0.5. Thus, in order to produce a flat distribution of mass ratios, the dynamical biasing parameter should not be too large, and this is what the model infers ($\alpha=0.6\pm1.0$).

In Fig. \ref{FIG:MARKOVDOTS} we plot those values of ${\cal Q}$ generated along the Markov Chain that exceed $-1$ (i.e. those models that deliver output parameters that are collectively within $1\sigma$ of the observations), against the different model input parameters. These plots show that the best-fit model input parameters are all well defined, apart from $\mu_{_{\rm C}}$. Fortunately $\mu_{_{\rm C}}$ is already quite well constrained by observation, and likely to become better constrained in the future. If $\mu_{_{\rm C}}$ were increased, the efficiency, $\eta$, would have to be reduced proportionately (or each core would have to produce more than one long-lived binary) --- and {\it vice versa}.

Fig. \ref{FIG:MONTAGE} illustrates how ${\cal Q}$ varies on planes through the best-fit solution, i.e. if just two of the model input parameters are varied. These plots are generated with a regular two-dimensional grid of models, and $10^7$ stars per model. On each row, the ordinate is the same for all five plots, and the abscissa cycles through the remaining five input parameters. From the plots in the first row we see that $\mu_{_{\rm C}}$ is weakly constrained, and also, from the second plot along this row, that if $\mu_{_{\rm C}}$ is increased, $\eta$ must be reduced proportionately. In all other cases, an horizontal scan of the plots in a row reveals that the parameter concerned (the ordinate) is very well, and uniquely, constrained by the observations.

\begin{figure}
\begin{center}
\includegraphics[width=0.48\textwidth,angle=0]{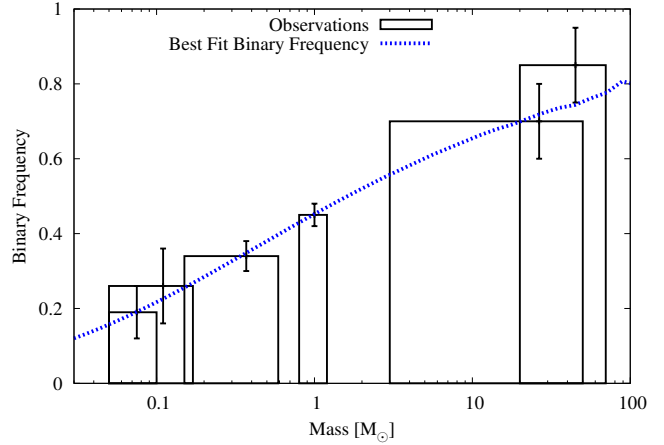}
\caption{The boxes represent the observational estimates of multiplicity frequency in different primary-mass intervals, as detailed in the text, and summarised in Table \ref{TAB:STELSTAT}. The error bars represent the observational uncertainties. The dashed line shows the multiplicity frequency as a function of primary mass for the best-fit model. The unruly points at large $M_1$ are due to small-number statistics.}
\label{FIG:BINFREQ}
\end{center}
\end{figure}

Fig. \ref{FIG:BINFREQ} presents the binary frequency as a function of primary mass, for the best-fit model, generated using $10^7$ stars, along with the observational data used to constrain the model. We reiterate that we do not use the two higher-mass points, only the four lower-mass points. Notwithstanding, the model fits all six points well.

\begin{figure}
\begin{center}
\includegraphics[width=0.48\textwidth,angle=0]{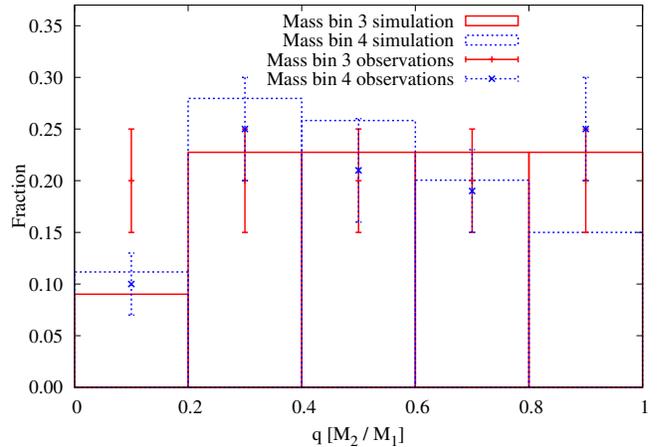}
\caption{The distribution of mass ratios for binaries having primaries in mass-bins 3 and 4. The plotted symbols with error bars represent the observationally inferred expectation values and uncertainties: orange plus signs for mass-bin 3, and blue crosses for mass-bin 4. The histograms represent the model results: orange solid line for mass-bin 3, and dotted blue line for mass-bin 4.}
\label{FIG:MASSRATIO}
\end{center}
\end{figure}

Fig. \ref{FIG:MASSRATIO} presents the mass ratio distributions for binaries having primaries in mass-bins 3 and 4. We see that there is acceptable agreement. The largest divergence occurs in the extreme bins. This is not surprising, given that, in the model, the components of a binary system are drawn from a log-normal distribution of masses with a power-law weighting.

\section{Discussion}\label{SEC:DIS}%

\subsection{Critique of the model}%

The critical assumption of the model is that each core spawns, on average, exactly one long-lived binary system, i.e. one binary system that survives to populate the field. If this assumption were relaxed, in the sense that a core might spawn more than one long-lived binary system (say, on average ${\cal B}$ binary systems), then $\eta$ and ${\cal N}_{_{\rm O}}$ would have to be increased (in proportion to ${\cal B}$). Conversely, if not all cores were to spawn a binary system, $\eta$ and ${\cal N}_{_{\rm O}}$ would have to be reduced, but it would then becomes impossible to reproduce the variation of binary frequency with primary mass, $b(M_{_1})\;$ ---$\;$ unless one were to introduce an additional parameter to allow the efficiency to be much higher for cores that spawn binaries than for those that don't.

If the observed estimate for the overall binary frequency of low-mass field stars (i.e. binaries with primaries in the range $(0.02,2.0)\,{\rm M}_{_\odot}$) were to increase, this would reduce ${\cal N}_{_{\rm O}}$, and consequently $\eta$. For example, if the observed  overall binary frequency of low-mass field stars were increased to $0.5$, the model would require ${\cal N}_{_{\rm O}}\!\sim\!3$ and $\eta\!\sim\!0.8\pm0.2$.

It is difficult to see how the various standard deviations could change much, unless $\sigma_{_{\rm S}}$ is very different from the \citet{Chabrier2005} value. If $\sigma_{_{\rm S}}$ were larger, $\sigma_{_{\rm C}}$ and $\sigma_{_{\rm O}}$ could also be larger, and {\it vice versa}.

If the distribution of mass ratios were skewed in favour of systems with comparable mass, i.e. $dp_q/dq>0$, then $\sigma_{_{\rm O}}$ would need to be reduced, and/or $\alpha$ increased (more dynamical biasing).

\subsection{Previous theoretical work}\label{SEC:OTHER}%


Some of the consequences of a self-similar mapping are explored by \citet{Clarke1996}, but with different distribution functions, and less emphasis on observational constraints. 

\citet{Swift+Williams2008} develop a similar model to ours, but one which includes a power-law extension to the CMF at high masses, based on the analysis of \citet{Padoan+NordlundCMF2002}, and which invokes somewhat different model parameters. They explore the consequences of varying the prescriptions for generating multiple systems, and for sub-fragmentation of a core, but their work differs from ours in that they do not explore in depth the question of multiplicity and its variation with primary mass, and they do not draw any firm conclusions on the efficiency, or on the number of stars spawned by a single core.

\citet{GoodwinetalCMF2IMF2008} explore the consequences of multiplicity for the mapping from the CMF into the IMF, and in particular the effect of multiplicity on the extremes of the IMF. Their preferred model presumes that all cores spawn multiple systems, with the number of stars in a system increasing very slightly with the mass of the progenitor core (the model is therefore not strictly self-similar), and it has quite a low efficiency, $\eta_{_{\rm O}}=0.27$. They do not explore the issue of how such systems might subsequently evolve to produce singles, so they cannnot exploit the observed variation of binary fraction with primary mass.

\citet{Goodwin+KouwenhovenCMF2IMF2009} demonstrate that the mapping from a log-normal CMF into an approximately log-normal System Mass Function (SMF) and from the SMF into an approximately log-normal IMF admits a wide range of prescriptions for (i) how the efficiency, varies with the core mass ($\eta(M_{_{\rm C}})$), (ii) whether the probability that a core spawns a single or a binary depends on its mass (effectively  ${\cal N}(M_{_{\rm C}})$), (iii) and the distribution of mass ratios in such binaries. This concurs with our conclusion (see Section \ref{SEC:ADDPARAMS}) that, whilst there are many theoretical arguments for allowing the input parameters of the mapping to depend on the core mass (thereby rendering the mapping non self-similar), the effect on the IMF is so subtle that these dependencies cannot usefully be constrained by the existing observations.

\subsection{Additional model parameters}\label{SEC:ADDPARAMS}%

We have considered the following refinements to the model. However, none of them is justified, since none of them, either individually or in combination, produces a significant improvement to the fit; in respect of items (ii), (iii) and (iv), \citet{Goodwin+KouwenhovenCMF2IMF2009} reached essentially the same conclusion, but on the basis of a very different model and less restrictive observational constraints. Necessarily, all these refinements would corrupt the self-similarity of the mapping.

\begin{enumerate}

\item{We have explored models in which the lifetime of a prestellar core (i.e. the time during which a prestellar core is detected as such) depends on its mass according to $t_{_{\rm C}}\propto M_{_{\rm C}}^{\,\chi_t}$. Negative values of $\chi_t$ skew the IMF towards high masses, because low-mass cores are over-represented in the CMF. Conversely, positive values of $\chi_t$ skew the model IMF towards low masses, because high-mass cores are over-represented in the CMF. There is no consensus on this. \citet{Hatchell+Fuller2008} have argued that more massive cores evolve faster than less massive ones, and are therefore under-represented in the CMF; this might be taken into the reckoning with $\chi_t\,=-\,0.25$. Conversely, \citet*{Clark+Klessen+Bonnell2007} have argued that massive cores, being more diffuse have longer lifetimes, and are therefore over-represented in the CMF; on the basis of a simple free-fall argument, and Larson's scaling relations, this might be taken into the reckoning with $\chi_t\,=+\,0.25$.}

\item{We have explored models in which the efficiency of star formation in a prestellar core depends on its mass according to $\eta_{_{\rm O}}\propto M_{_{\rm C}}^{\chi_\eta}$. This is equivalent to including feedback from massive stars. Star formation is promoted by feedback from massive stars if $\chi_\eta$ is positive, and suppressed if $\chi_\eta$ is negative. However, it is not known what the sense of feedback from massive stars is, on the scale of a single core.}

\item{We have explored models in which the number of stars formed from a prestellar core depends on its mass according to ${\cal N}_{_{\rm O}}\propto M_{_{\rm C}}^{\,\chi_{_{\cal N}}}$. Negative values of $\chi_{_{\cal N}}$ (a) skew the IMF towards high masses, and (b) increase the multiplicity frequency of high-mass stars and reduce the multiplicity frequency of low-mass stars. Positive values of $\chi_{_{\cal N}}$ have the opposite effects. It is difficult to believe that  ${\cal N}$ does not increase with core mass (positive $\chi_{_{\!\cal N}}$). However, this would completely undermine the original argument for a self-similar mapping between CMF and IMF, namely that the high-mass slopes of the CMF and IMF appear to be indistinguishable. Moreover, in practice, the observational constraints can more easily accommodate the effects of negative $\chi_{_{\cal N}}$. Either way, non-zero $\chi_{_{\!\cal N}}$-values are not actually needed to fit the observational constraints we have invoked..}

\item{We have explored models in which the logarithmic range of stellar masses formed from a prestellar core depends on its mass according to $\sigma_{_{\rm O}}\propto M_{_{\rm C}}^{\chi_\sigma}$. It is probably the case that only positive values of $\chi_\sigma$ could be justified (i.e. higher-mass cores spawning a greater logarithmic spread of stellar masses), but this is not needed to fit the observational constraints. Moreover, it suppresses the high-mass end of the IMF, which -- in this purely log-normal model -- is already too low.}

\item{We have explored the possibility that there is some variance in, for example, ${\cal N}_{_{\rm O}}$, so that when ${\cal N}_{_{\rm O}}=3$ (say) not all cores spawn exactly three stars. However, firstly this introduces an additional model parameter, which should be avoided if possible, and secondly it makes no significant difference to the results, unless the variance is extremely large, so we do not include it in the basic model.}

\end{enumerate}

We reiterate that we are not arguing that these additional effects do not occur in nature. We are simply pointing out (a) that they are not justified by the currently available observational constraints, that is, one can obtain a good fit to the observations without them; and (b) that they would corrupt a self-similar mapping.

\section{Conclusions}\label{SEC:CON}%

We have developed a simple model to describe the mapping of the CMF onto the IMF.

\begin{itemize}

\item{The model has four assumptions: the central portions of the CMF and IMF are both log-normal; the mapping from the CMF onto the IMF is self-similar; if a core forms more than one star two of the stars end up in a long-lived binary; and the probability of a star of mass $M$ being in this binary is proportional to $M^\alpha$.}

\item{The model has six input parameters: $\mu_{_{\rm C}}$ and $\sigma_{_{\rm C}}$ are the logarithmic mean and standard deviation of the log-normal CMF; $\eta_{_{\rm O}}$ is the efficiency (i.e. the fraction of a core's mass that ends up in new stars); ${\cal N}_{_{\rm O}}$ is the mean number of stars spawned by a single core; $\sigma_{_{\rm O}}$ is the standard deviation of the log-normal distribution of relative stellar masses spawned by a single core; and $\alpha$ is the dynamical biasing parameter.}

\item{This model is able to fit the observed IMF, the observed binary frequency as a function of primary mass, and the observed distributions of mass ratio for binaries having Sun-like and M-dwarf primaries. The best fit requires $\mu_{_{\rm C}}=-0.03\pm0.10,\;\sigma_{_{\rm C}}=0.47\pm0.04,\;\eta=1.01\pm0.27,\;{\cal N}_{_{\rm O}}=4.34\pm0.43,\;\sigma_{_{\rm O}}=0.30\pm0.03,\;$ and $\alpha=0.87\pm0.64\,$. It fits the observations to within $0.25\sigma$.}

\end{itemize}

We have not demonstrated, nor do we advocate, that the mapping is necessarily self-similar, only that, if one assumes self-similarity, there is a simple mapping that fits the observational constraints well and therefore -- on the basis of Occam's Razor -- should be given consideration.

Moreover, if the mapping is not (at least, approximately) self-similar, then the notion that the shape of the IMF is inherited from the CMF must be abandoned.

Either way, there is a question to be answered beyond understaning the origin of the CMF: {\it either} why is the mapping self-similar, {\it or} why does the mapping, despite not being self-similar, produce an IMF with the same shape as the CMF?

The self-similar model suggests that the efficiency of star formation within a prestellar core is significantly higher ($\eta_{_{\rm O}}\simeq 1.0\pm 0.3$) than has previously been proposed \citep[e.g. $\eta_{_{\rm O}}\sim 0.3$, ][]{Alvesetal2007}. It also suggests that most stars, including singles, are born in small groups of $\sim 4$. This contrasts with the conclusion of \citet{LadaC2006} that most stars, being single, are born in isolation. Interestingly \citet{Nakamura2012} have recently reported evidence that prestellar cores are more fragmentated than had previously been thought. If cores spawn many stars, we may see multiple outflows from some cores \citep[e.g.][]{WuPetal2006}, but these outflows do not have to disperse a large fraction of the core's initial mass, and can simply punch holes in the residual envelope.

\section*{Acknowledgments}

We thank Cathie Clarke, Thijs Kouwenhoven, Mike Meyers and Peter Coles for useful discussions that helped to improve this paper. We gratefully acknowledge the support of the UK STFC, via a doctoral training account (KH) and a rolling grant (APW; PP/E000967/1). SKW gratefully acknowledges the support of the DFG Priority Programme No. 1573. SKW, SPG and APW gratefully acknowledge the support of the Marie Curie CONSTELLATION Research Training Network.

\bibliography{AntsBiblio} %

\begin{thebibliography}{55}
\expandafter\ifx\csname natexlab\endcsname\relax\def\natexlab#1{#1}\fi

\bibitem[{{Alves} {et~al}\mbox{.}(2007){Alves}, {Lombardi}, \&
  {Lada}}]{Alvesetal2007}
{Alves} J., {Lombardi} M., {Lada} C.~J., 2007, \aap, 462, L17

\bibitem[{{Basri} \& {Reiners}(2006)}]{Basri+Reiners2006}
{Basri} G., {Reiners} A., 2006, \aj, 132, 663

\bibitem[{{Bastian} {et~al}\mbox{.}(2010){Bastian}, {Covey}, \&
  {Meyer}}]{BastianNetal2010}
{Bastian} N., {Covey} K.~R., {Meyer} M.~R., 2010, \araa, 48, 339

\bibitem[{{Bate}(2012)}]{Bate2012}
{Bate} M.~R., 2012, \mnras, 419, 3115

\bibitem[{{Bressert} {et~al}\mbox{.}(2010){Bressert}, {Bastian}, {Gutermuth},
  {Megeath}, {Allen}, {Evans}, {Rebull}, {Hatchell}, {Johnstone}, {Bourke},
  {Cieza}, {Harvey}, {Merin}, {Ray}, \& {Tothill}}]{Bressertetal10}
{Bressert} E. {et~al.}, 2010, \mnras, 409, L54

\bibitem[{{Chabrier}(2003)}]{ChabrierStIMF2003}
{Chabrier} G., 2003, \pasp, 115, 763

\bibitem[{{Chabrier}(2005)}]{Chabrier2005}
{Chabrier} G., 2005, in Astrophysics and Space Science Library, Vol. 327, The
  Initial Mass Function 50 Years Later, {E.~Corbelli, F.~Palla, \&
  H.~Zinnecker}, ed., pp. 41--50

\bibitem[{{Chen} {et~al}\mbox{.}(2013){Chen}, {Arce}, {Zhang}, {Bourke},
  {Launhardt}, {Jorgensen}, {Lee}, {Foster}, {Dunham}, {Pineda}, \&
  {Henning}}]{Chenetal2013}
{Chen} X. {et~al.}, 2013, ArXiv e-prints

\bibitem[{{Clark} {et~al}\mbox{.}(2007){Clark}, {Klessen}, \&
  {Bonnell}}]{Clark+Klessen+Bonnell2007}
{Clark} P.~C., {Klessen} R.~S., {Bonnell} I.~A., 2007, \mnras, 379, 57

\bibitem[{{Clarke}(1996)}]{Clarke1996}
{Clarke} C.~J., 1996, \mnras, 283, 353

\bibitem[{{Close} {et~al}\mbox{.}(2003){Close}, {Siegler}, {Freed}, \&
  {Biller}}]{Closeetal2003}
{Close} L.~M., {Siegler} N., {Freed} M., {Biller} B., 2003, \apj, 587, 407

\bibitem[{{Elmegreen} {et~al}\mbox{.}(2008){Elmegreen}, {Klessen}, \&
  {Wilson}}]{Elmegreen2008}
{Elmegreen} B.~G., {Klessen} R.~S., {Wilson} C.~D., 2008, \apj, 681, 365

\bibitem[{{Enoch} {et~al}\mbox{.}(2008){Enoch}, {Evans}, {Sargent}, {Glenn},
  {Rosolowsky}, \& {Myers}}]{Enochetal2008}
{Enoch} M.~L., {Evans}, II N.~J., {Sargent} A.~I., {Glenn} J., {Rosolowsky} E.,
  {Myers} P., 2008, \apj, 684, 1240

\bibitem[{{Enoch} {et~al}\mbox{.}(2006){Enoch}, {Young}, {Glenn}, {Evans},
  {Golwala}, {Sargent}, {Harvey}, {Aguirre}, {Goldin}, {Haig}, {Huard},
  {Lange}, {Laurent}, {Maloney}, {Mauskopf}, {Rossinot}, \&
  {Sayers}}]{Enochetal2006}
{Enoch} M.~L. {et~al.}, 2006, \apj, 638, 293

\bibitem[{{Girichidis} {et~al}\mbox{.}(2012){Girichidis}, {Federrath},
  {Banerjee}, \& {Klessen}}]{Girichidis2012}
{Girichidis} P., {Federrath} C., {Banerjee} R., {Klessen} R.~S., 2012, \mnras,
  420, 613

\bibitem[{{Goodwin} \& {Kouwenhoven}(2009)}]{Goodwin+KouwenhovenCMF2IMF2009}
{Goodwin} S.~P., {Kouwenhoven} M.~B.~N., 2009, \mnras, 397, L36

\bibitem[{{Goodwin} {et~al}\mbox{.}(2008){Goodwin}, {Nutter}, {Kroupa},
  {Ward-Thompson}, \& {Whitworth}}]{GoodwinetalCMF2IMF2008}
{Goodwin} S.~P., {Nutter} D., {Kroupa} P., {Ward-Thompson} D., {Whitworth}
  A.~P., 2008, \aap, 477, 823

\bibitem[{{Hatchell} \& {Fuller}(2008)}]{Hatchell+Fuller2008}
{Hatchell} J., {Fuller} G.~A., 2008, \aap, 482, 855

\bibitem[{{Hennebelle} \& {Chabrier}(2008)}]{Hennebelle+ChabrierCMF2008}
{Hennebelle} P., {Chabrier} G., 2008, \apj, 684, 395

\bibitem[{{Hennebelle} \& {Chabrier}(2009)}]{Hennebelle+ChabrierCMF2009}
{Hennebelle} P., {Chabrier} G., 2009, \apj, 702, 1428

\bibitem[{{Hubber} \& {Whitworth}(2005)}]{Hubber+Whitworth2005}
{Hubber} D.~A., {Whitworth} A.~P., 2005, \aap, 437, 113

\bibitem[{{Janson} {et~al}\mbox{.}(2012){Janson}, {Hormuth}, {Bergfors},
  {Brandner}, {Hippler}, {Daemgen}, {Kudryavtseva}, {Schmalzl}, {Schnupp}, \&
  {Henning}}]{Janson2012}
{Janson} M. {et~al.}, 2012, \apj, 754, 44

\bibitem[{{Johnstone} \& {Bally}(2006)}]{Johnstone+BallyCMF2006}
{Johnstone} D., {Bally} J., 2006, \apj, 653, 383

\bibitem[{{Johnstone} {et~al}\mbox{.}(2001){Johnstone}, {Fich}, {Mitchell}, \&
  {Moriarty-Schieven}}]{JohnstoneetalCMF2001}
{Johnstone} D., {Fich} M., {Mitchell} G.~F., {Moriarty-Schieven} G., 2001,
  \apj, 559, 307

\bibitem[{{Johnstone} {et~al}\mbox{.}(2000){Johnstone}, {Wilson},
  {Moriarty-Schieven}, {Joncas}, {Smith}, {Gregersen}, \&
  {Fich}}]{JohnstoneetalCMF2000}
{Johnstone} D., {Wilson} C.~D., {Moriarty-Schieven} G., {Joncas} G., {Smith}
  G., {Gregersen} E., {Fich} M., 2000, \apj, 545, 327

\bibitem[{{K{\"o}hler} {et~al}\mbox{.}(2008){K{\"o}hler}, {Neuh{\"a}user},
  {Kr{\"a}mer}, {Leinert}, {Ott}, \& {Eckart}}]{Kohler2008}
{K{\"o}hler} R., {Neuh{\"a}user} R., {Kr{\"a}mer} S., {Leinert} C., {Ott} T.,
  {Eckart} A., 2008, \aap, 488, 997

\bibitem[{{K{\"o}nyves} {et~al}\mbox{.}(2010){K{\"o}nyves}, {Andr{\'e}},
  {Men'shchikov}, {Schneider}, {Arzoumanian}, {Bontemps}, {Attard}, {Motte},
  {Didelon}, {Maury}, \& {et al.}}]{Veraetal2010}
{K{\"o}nyves} V. {et~al.}, 2010, \aap, 518, L106

\bibitem[{{Kroupa}(1995)}]{Kroupa1995}
{Kroupa} P., 1995, \mnras, 277, 1491

\bibitem[{{Kroupa}(2001)}]{KroupaStIMF2001}
{Kroupa} P., 2001, \mnras, 322, 231

\bibitem[{{Lada}(2006)}]{LadaC2006}
{Lada} C.~J., 2006, \apjl, 640, L63

\bibitem[{{Mason} {et~al}\mbox{.}(1998){Mason}, {Gies}, {Hartkopf}, {Bagnuolo},
  {ten Brummelaar}, \& {McAlister}}]{MasonBetal1998}
{Mason} B.~D., {Gies} D.~R., {Hartkopf} W.~I., {Bagnuolo}, Jr. W.~G., {ten
  Brummelaar} T., {McAlister} H.~A., 1998, \aj, 115, 821

\bibitem[{{Matzner} \& {McKee}(2000)}]{Matzner+McKee2000}
{Matzner} C.~D., {McKee} C.~F., 2000, \apj, 545, 364

\bibitem[{{McDonald} \& {Clarke}(1993)}]{McDonald+Clarke93}
{McDonald} J.~M., {Clarke} C.~J., 1993, \mnras, 262, 800

\bibitem[{{McDonald} \& {Clarke}(1995)}]{McDonald+Clarke95}
{McDonald} J.~M., {Clarke} C.~J., 1995, \mnras, 275, 671

\bibitem[{{Motte} {et~al}\mbox{.}(1998){Motte}, {Andre}, \&
  {Neri}}]{MotteetalCMF1998}
{Motte} F., {Andre} P., {Neri} R., 1998, \aap, 336, 150

\bibitem[{{Motte} {et~al}\mbox{.}(2001){Motte}, {Andr{\'e}}, {Ward-Thompson},
  \& {Bontemps}}]{MotteetalCMF2001}
{Motte} F., {Andr{\'e}} P., {Ward-Thompson} D., {Bontemps} S., 2001, \aap, 372,
  L41

\bibitem[{{Nakamura} {et~al}\mbox{.}(2012){Nakamura}, {Takakuwa}, \&
  {Kawabe}}]{Nakamura2012}
{Nakamura} F., {Takakuwa} S., {Kawabe} R., 2012, \apjl, 758, L25

\bibitem[{{Nutter} \& {Ward-Thompson}(2007)}]{Nutter+WardTCMF2007}
{Nutter} D., {Ward-Thompson} D., 2007, \mnras, 374, 1413

\bibitem[{{Padoan} \& {Nordlund}(2002)}]{Padoan+NordlundCMF2002}
{Padoan} P., {Nordlund} {\AA}., 2002, \apj, 576, 870

\bibitem[{{Padoan} {et~al}\mbox{.}(2007){Padoan}, {Nordlund}, {Kritsuk},
  {Norman}, \& {Li}}]{PadoanetalCMF2007}
{Padoan} P., {Nordlund} {\AA}., {Kritsuk} A.~G., {Norman} M.~L., {Li} P.~S.,
  2007, \apj, 661, 972

\bibitem[{{Preibisch} {et~al}\mbox{.}(1999){Preibisch}, {Balega}, {Hofmann},
  {Weigelt}, \& {Zinnecker}}]{Preibischetal1999}
{Preibisch} T., {Balega} Y., {Hofmann} K., {Weigelt} G., {Zinnecker} H., 1999,
  \na, 4, 531

\bibitem[{{Raghavan} {et~al}\mbox{.}(2010){Raghavan}, {McAlister}, {Henry},
  {Latham}, {Marcy}, {Mason}, {Gies}, {White}, \& {ten
  Brummelaar}}]{Raghavan2010}
{Raghavan} D. {et~al.}, 2010, \apjs, 190, 1

\bibitem[{{Rathborne} {et~al}\mbox{.}(2009){Rathborne}, {Lada}, {Muench},
  {Alves}, {Kainulainen}, \& {Lombardi}}]{RathborneetalCMF2009}
{Rathborne} J.~M., {Lada} C.~J., {Muench} A.~A., {Alves} J.~F., {Kainulainen}
  J., {Lombardi} M., 2009, \apj, 699, 742

\bibitem[{{Reggiani} \& {Meyer}(2011)}]{Reggiani2011}
{Reggiani} M.~M., {Meyer} M.~R., 2011, \apj, 738, 60

\bibitem[{{Reipurth} \& {Zinnecker}(1993)}]{Reipurth+Zinnecker1993}
{Reipurth} B., {Zinnecker} H., 1993, \aap, 278, 81

\bibitem[{{Simpson} {et~al}\mbox{.}(2008){Simpson}, {Nutter}, \&
  {Ward-Thompson}}]{SimpsonetalCMF2008}
{Simpson} R.~J., {Nutter} D., {Ward-Thompson} D., 2008, \mnras, 391, 205

\bibitem[{{Smith} {et~al}\mbox{.}(2011){Smith}, {Glover}, {Bonnell}, {Clark},
  \& {Klessen}}]{SmithR2011}
{Smith} R.~J., {Glover} S.~C.~O., {Bonnell} I.~A., {Clark} P.~C., {Klessen}
  R.~S., 2011, \mnras, 411, 1354

\bibitem[{{Stanke} {et~al}\mbox{.}(2006){Stanke}, {Smith}, {Gredel}, \&
  {Khanzadyan}}]{StankeetalCMF2006}
{Stanke} T., {Smith} M.~D., {Gredel} R., {Khanzadyan} T., 2006, \aap, 447, 609

\bibitem[{{Sterzik} \& {Durisen}(1998)}]{Sterzik+Durisen98}
{Sterzik} M.~F., {Durisen} R.~H., 1998, \aap, 339, 95

\bibitem[{{Swift} \& {Williams}(2008)}]{Swift+Williams2008}
{Swift} J.~J., {Williams} J.~P., 2008, \apj, 679, 552

\bibitem[{{Testi} \& {Sargent}(1998)}]{Testi+SargentCMF1998}
{Testi} L., {Sargent} A.~I., 1998, \apjl, 508, L91

\bibitem[{{van Albada}(1968{\natexlab{a}})}]{VanAlbada68a}
{van Albada} T.~S., 1968{\natexlab{a}}, \bain, 19, 479

\bibitem[{{van Albada}(1968{\natexlab{b}})}]{VanAlbada68b}
{van Albada} T.~S., 1968{\natexlab{b}}, \bain, 20, 57

\bibitem[{{Wu} {et~al}\mbox{.}(2009){Wu}, {Takakuwa}, \& {Lim}}]{WuPetal2006}
{Wu} P.-F., {Takakuwa} S., {Lim} J., 2009, \apj, 698, 184

\bibitem[{{Young} {et~al}\mbox{.}(2006){Young}, {Enoch}, {Evans}, {Glenn},
  {Sargent}, {Huard}, {Aguirre}, {Golwala}, {Haig}, {Harvey}, {Laurent},
  {Mauskopf}, \& {Sayers}}]{YoungetalCMF2006}
{Young} K.~E. {et~al.}, 2006, \apj, 644, 326

\end{thebibliography}

\bsp

\label{lastpage}

\end{document}